\documentclass[usenatbib]{mnras}
\usepackage[T1]{fontenc}
\usepackage{ae,aecompl}

\usepackage{graphicx}
\usepackage{amsmath}
\usepackage{amssymb}

\usepackage{cuted}
\usepackage{multirow}
\usepackage[version=4]{mhchem}

\title[\ce{CO2} planetary atmospheres and stellar environment]{A systematic study of \ce{CO2} planetary atmospheres and their link to the stellar environment}

\author[A. Petralia et al.]{A. Petralia$^{1}$ \hspace{-0.2cm}\thanks{E-mail: antonino.petralia@inaf.it}, E. Alei$^{2}$ \hspace{-0.2cm}\thanks{ETH Z{\"u}rich, Institute for Particle Physics and Astrophysics, Wolfgang-Pauli-Str. 27, 8093 Z{\"u}rich, Switzerland}, G. Aresu$^{3}$, D. Locci$^{1}$, C. Cecchi-Pestellini$^{1}$, G. Micela$^{1}$, \cr R. Claudi$^{2}$ and A. Ciaravella$^{1}$
\newauthor{}
\\
$^{1}$INAF - Osservatorio Astronomico di Palermo, P.za Parlamento 1, 90134 Palermo, Italy\\
$^{2}$INAF - Osservatorio Astronomico di Padova, Vicolo dell'Osservatorio 5, 35122 Padova, Italy\\
$^{3}$INAF - Osservatorio Astronomico di Cagliari, Via della Scienza 5, 09047 Selargius, Italy \\
}
\date{Accepted . Received ; in original form}
\pubyear{2019}
\begin{document}
\label{firstpage}
\pagerange{\pageref{firstpage}--\pageref{lastpage}}
\maketitle

\begin{abstract}
The Milky Way Galaxy is literally teeming with exoplanets; thousands of planets have been discovered, with thousands more planet candidates identified. Terrestrial-like planets are quite common around other stars, and are expected to be detected in large numbers in the future. Such planets are the primary targets in the search for potentially habitable conditions outside the solar system.

Determining the atmospheric composition of exoplanets is mandatory to understand their origin and evolution, as atmospheric processes play crucial roles in many aspects of planetary architecture. In this work we construct and exploit a 1D radiative transfer model based on the discrete-ordinates method in plane-parallel geometry. Radiative results are linked to a convective flux that redistributes energy at any altitude producing atmospheric profiles in radiative-convective equilibrium. The model has been applied to a large number (6250) of closely dry synthetic \ce{CO2} atmospheres, and the resulting pressure and thermal profiles have been interpreted in terms of parameter variability. Although less accurate than 3D general circulation models, not properly accounting for e.g., clouds and atmospheric and ocean dynamics, 1D descriptions are computationally inexpensive and retain significant value by allowing multidimensional parameter sweeps with relative ease.
\end{abstract}

\begin{keywords}
planets and satellites: atmospheres -- planets and satellites: terrestrial planets -- radiative transfer -- methods: numerical
\end{keywords}

\section{Introduction}
The past two decades have seen a major improvement in our knowledge of exoplanet properties, revealing an astonishing diversity in planetary masses, radii, mean temperatures, orbital parameters, and host stellar characteristics and influence. In particular rocky, terrestrial exoplanets such as  the one orbiting around Proxima Centauri \citep{Ang16} are found at an ever growing rate. It is estimated that approximately 30\% of stars in the solar neighborhood have planets with sizes within two times the Earth's radius and orbital periods within 85 days \citep{F13}. Around M~stars radial velocity surveys provide results even more remarkable: the occurrence of super Earths in the habitable zone  of these stars is about 40\% \citep{Bon13,Kop13,May14}. Thus, small planets orbiting small and cold stars seem to be quite ubiquitous within the Galaxy \citep{H13}.

An important issue is the potential habitability of terrestrial planets orbiting low-mass stars and cool dwarfs. Low-mass stars offer the opportunity for detecting and characterizing habitable terrestrial planets in the next decade. The conventional definition of habitable zone of a planet is usually limited to surface habitability, and the range of orbital distances within which suitable planets can maintain liquid water on their surfaces. Staying in the habitable zone does not clearly constitute a sufficient condition for a planet to preserve liquid water on its surface for geological times. Thus, the habitability of a planet relies on a complex array of geophysical and astrophysical factors. 

There is an extensive body of literature on the potential habitability of terrestrial planets in the habitable zone, using both 1D and 3D circulation models, based on Radiative Transfer (RT) platforms simulating synchronous and slow (e.g., \citealt{K16}), and fast rotating (e.g., \citealt{WT15}) planets. The difficulty of assessing the habitability of a planet in alien environments depends mostly on the reliability of simulations in reproducing inherently 3D processes, having only a rough estimate of the planet's past irradiation, and orbital history. 1D models are less performing than 3D simulations in capturing the mean vertical structure of planetary atmospheres, in particular in the case of terrestrial planets in the habitable zone of low-mass stars, and are generally less sensitive than higher dimensional models. Nevertheless, they are very useful to test the effects of some processes, or to explore a parameter space too broad for heavy 3D modeling, such as e.g., the impact of differences in the water vapor content (even in clear-sky conditions) on the inner edge of the habitability zone \citep{Y16}. In other studies, 1D models are used to study the validity of $k-$distribution when the stellar spectrum is correlated with the atmospheric absorption (H$_2$O also shapes the spectrum of cool stars), the effect of different collision-induced absorption parametrizations, foreign broadening (i.e. broadening induced by collision with molecules of other species), and various isotopologue ratios. In general 1D models rely just on a few  assumptions, and can be easily exploited in a large parameter space, being thus effective in  characterizing extrasolar planets.

The minimal set of observables to be derived by modeling efforts are pressure and temperature profiles, abundances and chemistry, and atmospheric dynamics. Even in such limited configuration, the model complexity must increase considerably, if models should reasonably represent the real world. For rocky planets the number of unknowns goes up because of uncertain initial conditions, the interactions between the surface, the interior and the atmosphere, and the possible impact of biological processes. In the simple case of 1D models, one possibility is to consider the energy balance along single air vertical columns, as implemented e.g., in Radiative-Convective (RC) models, in which the only dimension is the altitude.

The introduction of RC models in the 1960's was pioneering for understanding some of the first-order details of planetary climate and climate change (e.g., \citealt{MW67}). More recently, such models have been exploited to investigate exoplanet habitability either e.g., defining a general habitable zone for N$_2$/CO$_2$/H$_2$O planets {\citep{Kastetal93,Kop13}}, or focusing to single planets (e.g., Gliese 581d; {\citealt{vonParetal10,Woretal10,HD11}}). This class of models assumes that the vertical structure of an atmosphere is determined by the convection and radiation within the simulated atmospheric column. RC solutions to temperature profiles are computed numerically, frequently using the so-called convective adjustment. This last technique consists in simulating the effects of dry and/or moist convection by adjusting the lapse rate of temperature (and moisture) to specified profiles along the local atmospheric column. The lapse rate (either dry or moist) stabilizes the statistically unstable temperature profile derived from the radiative equilibrium in the deep layers of an atmosphere. Since the vertical temperature gradient changes the energy fluxes, the process must be iterated until the temperature and upwelling flux are continuous (see e.g., \citealt{MS64}). There is no explicit atmospheric circulation, although it is present implicitly, {through flux redistribution}. RC models may provide useful general indications of the impact on the planetary surface temperature of parameters such as {surface} pressure and CO$_2$ mixing ratio.

{Exploiting a RC model, we construct a large ensemble of 1D simulations (6,250) that systematically vary every relevant external parameter to represent possible exoplanet atmospheres, as we can reliably compute atmospheric profiles in a relatively short time. We want to relate variations} in vertical profiles and other atmospheric characteristics (such as e.g., the {surface} temperature) to chemical trends in the major constituents of the atmospheric gas, and investigate how these relations change with the incoming stellar radiation. These models are useful even if they do not correspond closely to reality, because they allow to confine strange and unfamiliar aspects of atmospheric processes. In Section~\ref{model} we briefly describe the {physical processes considered and the} techniques used in our atmospheric model. The model is validated in Section \ref{validation}. In Section~\ref{res} we confine the parameter space, and show the results. Last Section contains our conclusions.

\section{The model}\label{model}
The atmosphere is in hydrostatic balance, and divided into ${\cal N}_{\rm L}$ layers, based on pressure differences. The model consists conceptually of two {steps. Firstly, we solve the radiative transfer problem to generates a net radiative flux throughout the atmosphere, taking into account} absorption, scattering, and {thermal emission, i.e.} re-processing of the incoming stellar radiation. Simple radiative equilibrium gives results that are gravitationally unstable near the surface, and convection must take place to transport heat upward, reducing the rate at which the temperature in the atmosphere decreases (temperature lapse rate). {The second step consists, therefore,} in a convective flux that redistributes energy at any altitude.  

\subsection{The radiative transfer problem}
{In the solution to the RT problem, we describe the incident stellar flux as a Planck function at the effective temperature of the star, and consider the planetary surface to be Lambertian, i.e. having a grey isotropic reflectivity, $R_{\rm S}$. We consider clear sky conditions throughout the atmosphere. We do not also include emission from the top of the atmosphere (TOA), ignoring the existence of a layer of absorbing and re-emitting material.

We generate line by line absorption coefficients, and subsequently transform them through a $k-$distribution technique creating look up tables for the absorption coefficients of these species. To minimize the impact of interpolation errors in the radiative fluxes we precalculate the absorption coefficients also accounting for different CO$_2$ mixing ratios. The line by line absorption absoprtion coefficient is built as follows}
\newpage

\begin{strip}
\begin{equation}
\begin{split}
k_{\nu}=k_{\rm CO_2} + k_{\rm H_2O} + k_{\rm N_2} = \omega_{\rm CO_2} \left[ \sum_{\delta_{\nu_0} < 25\,\rm cm^{-1}} k_{\nu}^{\rm CO_2} +(1-\omega_{\rm CO_2})\left(\frac{p}{p_{\rm ref}}\right)\left(\frac{T_{\rm ref}}{T}\right)k_{f}^{\rm CO_2} +k_{\nu}^\star \right] + 
\\
+ \omega_{\rm H_2O} \left[ \sum_{\delta_{\nu_0} < 25\,\rm cm^{-1}} k_{\nu}^{\rm H_2O} +(1-\omega_{\rm H_2O})\left(\frac{p}{p_{\rm ref}}\right)\left(\frac{T_{\rm ref}}{T}\right)k^{\rm H_2O}_{f} +
\omega_{\rm H_2O}\left(\frac{p}{p_{\rm ref}}\right)\left(\frac{T_{\rm ref}}{T}\right)k^{\rm H_2O}_{s}\right]
+ \omega_{\rm N_2}  \sum_{\delta_{\nu_0} < 25\,\rm cm^{-1}}k_{\nu}^{\rm N_2}
\end{split}
\label{main_spec}
\end{equation}
\end{strip}
where $\omega_i$ is the Volume Mixing Ratio (VMR) of the $i-$th species, and $\delta_{\nu_0} = | \nu -\nu_0 |$; $k_\nu^\star$ indicates the CO$_2$ self collision induced absorption, and $f$ and $s$ stand for foreign- and self-continuum, respectively. We exploit the HITRAN2016 Molecular Spectroscopic Database \citep{G17} in the interval $0.24-1000~\mu$m ($10-42000$~cm$^{-1}$). The line {shapes} are Voigt {profiles}, opportunely modified for H$_2$O and CO$_2$ (see Appendix). 

Following \citet{Mis12} we divide the wavelength range $0.24-1000~\mu$m in 14 bands. Each band is then further partitioned in 10 sub-bands, {obtaining a total of 140 bands}, and this is the spacing where we apply the $k-$distribution technique. The externally stored $k-$distributed values {are defined in Table~\ref{kdisttab} and they are read and interpolated, according to the required values}.

\begin{table*}
\caption{Opacity table.}
\centering
\begin{tabular}{l l}
\hline
Parameter      &  Values \\ \hline 
Pressure (bar) &  $10^{-4}$, $10^{-3}$, $2.5\times10^{-3}$, $10^{-2}$, $2.5\times10^{-2}$, $10^{-1}$, $2.5\times10^{-1}$, $1$, $10$, $10^2$ \\
Temperature (K) &  $100$, $150$, $200$, $250$, $300$, $350$, $400$, $450$, $500$ \\
$\omega_{\rm CO_2}$ and $\omega_{\rm H_2O}$ &  $10^{-6}$, $10^{-5}$, $10^{-4}$, $10^{-3}$, $10^{-2}$, $10^{-1}$, $9.5\times10^{-1}$\\
\hline
\end{tabular}
\flushleft
\label{kdisttab}
\end{table*}

The $k-$distributed transmission is condensed in a few tens of points {(i.e. 32)} for each band, making the storage of absorption coefficients and their use for the RT calculation fast and simple. A second fundamental benefit of this method is that absorption coefficients at largely different temperature and pressure are now degraded at the same resolution, still maintaining the precise information on the value of the transmission. Since {$k-$distributed transmissions} are smooth, the RT problem is solved through Gaussian quadrature in each atmospheric layer from the TOA to the {surface}. 

{The scattering coefficients are calculated in the Rayleigh approximation as average of coefficients obtained for each molecular specie using the VMRs as weights}
\begin{equation}
k(\lambda) = 4.577\times10^{-21} \, \left( \frac{6+3D}{6-7D} \right) \frac{r^2}{\lambda^4}
\label{eq:SR}
\end{equation}
{where the numerical prefactor is taken from \citet{A73}, and the parameters $D$ and $r$ are the depolarization ratio and the refractivity, respectively. The numerical values of these last quantities are taken from the compilation of \citet[][and references therein]{VT07} in the case of molecular nitrogen and carbon dioxide; for water we use the prescriptions given in \cite{vonParetal10}.}

{The optical depths, obtained by considering both absorption and scattering, are given in input to the DISORT 4 module, the latest version of the DISORT (DIScrete Ordinate Radiative Transfer) software tool (see \citealt{Sta88}, for its first release), which solves the 1D RT problem by means of the discrete ordinate technique. We use 8 streams. Once the RT has been solved in each of the quadrature intervals, the flux in the layers is reconstructed using the same quadrature weights.}

{The temperature profile in the atmosphere is updated as follows}
\begin{equation}
\begin{split}
  T_{n} = T_{n-1} + \frac{{\rm d}T}{{\rm d} t}\Delta t  
  = T_{n-1} + \frac{g_{\rm P}}{c_P} \frac{{\rm \Delta}F}{{\rm \Delta} P} \Delta t   
\end{split}
\end{equation}
{where $T_{n-1}$ and $T_{n}$ are the temperature at the $(n-1)-$th and $n-$th time steps, $g_{\rm P}$ the planetary gravitational acceleration, $c_P$ the specific heat at fixed pressure of the atmospheric gas mixture, and ${\rm \Delta}F$ and ${\rm \Delta}P$  net flux and pressure differences between the top and the bottom of each atmospheric layer, respectively. The time step is chosen as the minimum between $16$~hrs and the one that would give a maximum temperature variation $\delta T=(g_{\rm P} / c_P) \times (\Delta F/\Delta P) \Delta t = 4$K.}

\subsection{Convective adjustment}\label{CA}
We simulate the convective redistribution of energy through convective adjustment, a procedure based on local mixing within unstable layers: whenever the radiative equilibrium lapse rate exceeds some pre-specified value marking the passage from the radiative to the convective regime (critical lapse rate, CLR), upward heat transport occurs to restore the lapse rate to that pre-specified value, while conserving the mean energy of the two layers \citep{MW67}. We use either constant and altitude dependent CLRs. In this latter case we exploit the following weighted expression for the CLR in a non saturated atmosphere
\begin{equation}
\Gamma_{\rm c} = -\left( \frac{{\rm d} T}{{\rm d} z} \right)_{\rm c} = \left(1-\phi \right) \Gamma_{\rm d} + \phi \Gamma_{\rm w}
\label{Gdry}
\end{equation}
where $z$ is the altitude, $\phi (z)$ being the relative humidity, and 
\begin{equation}
\Gamma_{\rm w} = \Gamma_{\rm d} \times \left( \frac{1+\omega_{\rm s} L/ {\cal R}_{\rm d} T}{1+\omega_{\rm s} L^2 /c_P {\cal R}_{\rm v} T^2}\right)
\label{gamma}
\end{equation}
{is the wet CLR in the adiabatic regime \citep{N09}. In equation (\ref{gamma}), $\Gamma_{\rm d}$ is the dry CLR in which we include \ce{CO2} condensation when required. Its value in the non-condensing regime is given by $\Gamma_{\rm d} = g_{\rm P}/c_p$. We use the same formalism as for water, with the exception of directly comparing pressures instead of defining a humidity. In the above expressions for the CLRs, equations (\ref{Gdry}) and (\ref{gamma})}, $L$ is the specific enthalpy of vaporisation, $\omega_{\rm s}$ the saturation vapor mixing ratio (the ratio between the mass of saturated vapor and the mass of dry air), and ${\cal R}_{\rm d}$ and ${\cal R}_{\rm v}$ the specific gas constants for dry air and vapor, respectively. We assume an initial isothermal profile at the atmospheric temperature of a leaky greenhouse model $T_{\rm s}= [1/(2-\epsilon)]^{1/4} T_{\rm eq}$, where $\epsilon$ is the {emissivity parameter at the TOA} and 
\begin{equation}
T_{\rm eq} = T_{\star} (1-A)^{1/4} \sqrt{\frac{R_{\star}}{2d_{\rm P}}} = \left( \frac {1-A}{4} \times \frac{S_\star} {\sigma}\right)^{1/4}
\label{teq}
\end{equation}
the equilibrium temperature of the planet. {In equation (\ref{teq}) $T_\star$ is the stellar temperature, $R_{\star}$ the star radius for a main sequence star \citep{Cox00}, $A$ the planetary albedo, $S_{\rm \star}$ the incident stellar flux at TOA, $\sigma$ the Stefan-Boltzmann constant,} and
\begin{equation}
d_{\rm P} = R_\star \sqrt{\frac{\sigma T_\star^4}{S_\star}}
\label{stpl-dist}
\end{equation}
the orbital distance. {In the determination of the initial temperature we set $A = 0$ in addition to $\epsilon$ = 0.}

In the case of wet atmospheres, we compute the initial water vapor pressure as $P_{\rm w} = \phi(z) \times P_{\rm sat}$, where $P_{\rm sat}$ is the saturation pressure. The humidity profile is $\phi (z) = 1.02~\phi_{\rm surf} \times ({\cal Q}(z) - 0.02)$ \citep{MW67}, where the factor ${\cal Q}(z)$ is the ratio between the initial pressure of the layer, $P(z)$ and the initial {surface} pressure $P_{\rm surf}$, {and $\phi_{\rm surf}$ is the humidity at the surface}. The minimum allowed value of this quantity is ${\cal Q} = 0.02$, and we forced any lower value to such limit, resulting in dry layers at low pressures (i.e. low ${\cal Q}$). Depending on the local temperature, the partial pressure of water is calculated at each iterative step, and thus the total pressure. 

\section{Validating the code} \label{validation}
\subsection{RT validation}
We consider an hypothetical early Mars atmosphere distributed in ${\cal N}_{\rm L} = 100$ layers predominantly of CO$_2$ composition (95\%) with a surface pressure $P_{\rm surf} = 500$~mbar. {Temperature and water VMR profiles  are taken from \citet{Mis12}}. The temperature rises rather linearly from a {surface} temperature $T_{\rm surf} = 250$~K to $\sim 170$~K at a pressure of 100~mbar, and then it is let constant. {The same occurs for water vapor whose atmospheric distribution follows closely the temperature profile}. At the {surface} $\omega_{\rm \ce{H2O}} = 1.5 \times 10^{-3}$, and it decreases upwards to $\omega_{\rm \ce{H2O}} = 10^{-7}$ at 100 mbar, where it remains constant for lower pressures. The remaining $\sim 5\%$ atmospheric gas is N$_2$. {To compare our results with others obtained using similar models we call the portion of spectrum between 0.24 and 4.6~$\mu$m \emph{solar}, while the remaining part extending to the far infrared \emph{IR}. This separation is artificial, and  it does not play any role in the RT solution. It will not be used elsewhere in this work.}

\begin{figure}
\centering
\includegraphics[width=8cm]{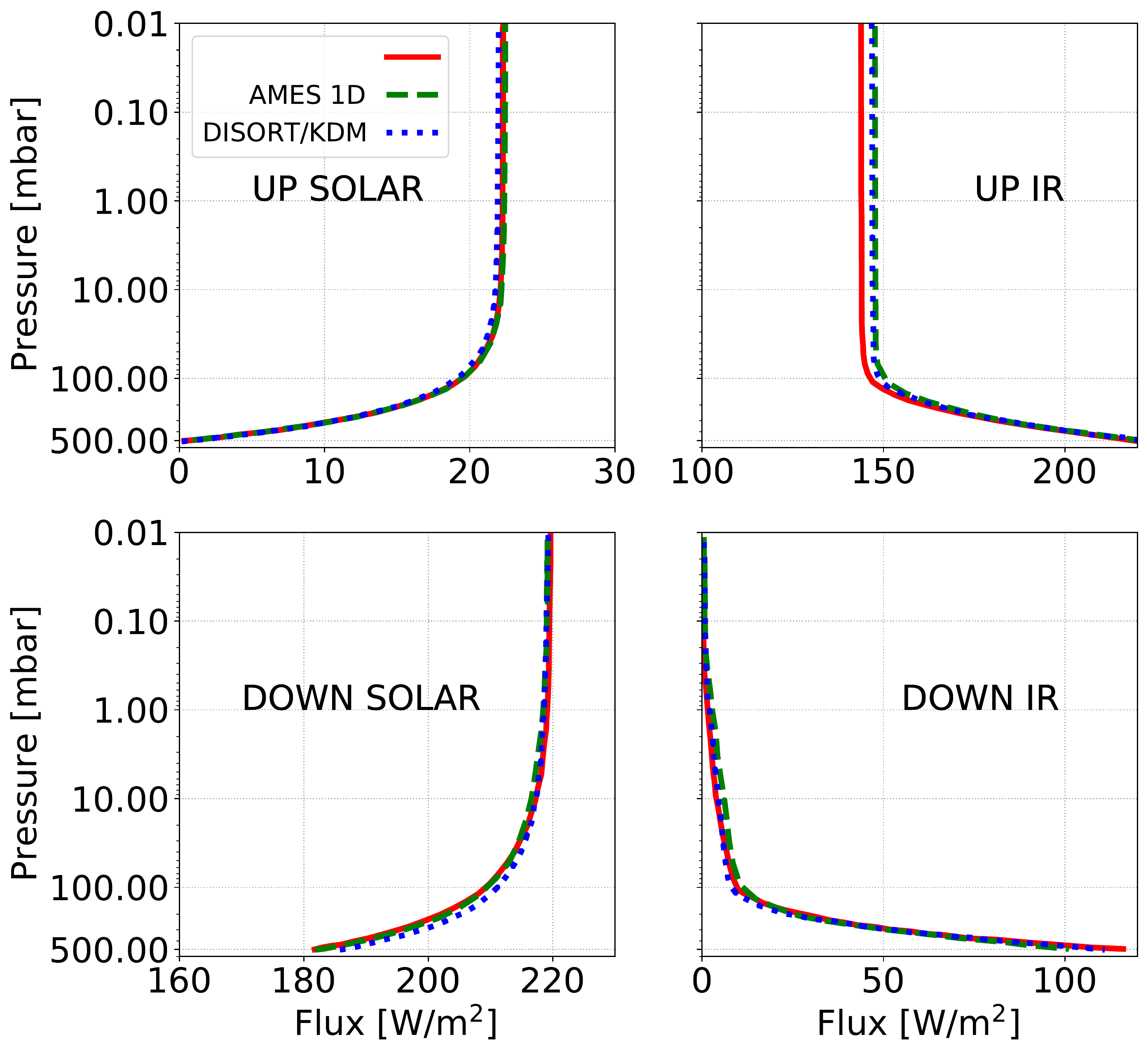}
\caption{Upward and downward fluxes in the solar and IR bands. Our RT model: red solid lines; NASA Ames model: green dashed lines; \citet{Mis12} absorption coefficients incorporated in our RT code: blue dotted lines.}
\label{fone}
\end{figure}

We set the {surface reflectivity $R_{\rm S} = 0$}, and we consider an irradiation perpendicular to the horizontal plane, i.e. cos$(\varphi) = 1$, $\varphi$ being the irradiation Solar (Stellar) Zenith Angle (SZA). In this section we do not include \ce{CO2} condensation in the model, not even as a simple correction to the $P-T$ profile {(e.g., \citealt{HD11,O16})}. 

In Figure~\ref{fone} we plot the upward and downward fluxes in the solar and IR bands. We set the solar radiation intensity equal to 220~W~m$^{-2}$, which is approximately the expected illumination for a young Mars ($\sim 75\%$ of the current solar intensity). This value also includes a factor $0.5$ {for slowly rotating planets}. To validate our code we compare the present results with those obtained using a similar 1D $k-$distribution RT model developed at NASA Ames \citep{Too89}. We also incorporate the absorption coefficients calculated from \citet{Mis12} into our RT module, and show the resulting outcomes in the figure. The different cross sections and the different RT module do not introduce important differences in the downward fluxes, while there is $\sim 2.5$~W~m$^{-2}$ difference in the upward solar flux, where Rayleigh scattering dominates. This is a straightforward consequence of the different resolutions at which calculations are performed by the different models. Contrary to our model, in both NASA Ames and \citet{Mis12} model opacities the Rayleigh scattering cross section in a given spectral range is calculated at the center of the band. Because the Rayleigh scattering depends steeply on wavelengths, low resolution calculations tend to underestimate the cross sections.

\subsection{Testing the RC model}
Once validated the RT model, we apply the integrated model (RT + RC descriptions) to a present-day Mars-like, synthetic atmosphere. The gas is predominantly composed by \ce{CO2} (95\%) with a mean surface pressure $P_{\rm surf} = 6.72$~mbar. \ce{N2} fills the remaining 5\% of the atmospheric gas. We assume a completely dry atmosphere. The parameters describing the relevant conditions in the atmosphere, and the boundary conditions are reported in Table~\ref{ttone}.

\begin{table}
\caption{Calculation parameters for the present-day Mars simulation.}
\centering
\begin{tabular}{l c c}
\hline
Parameter      &                           & Value \\ \hline 
Stellar irradiation & $S_\star$ (Wm$^{-2}$)& 586  \\
Stellar temperature & $T_\star$ (K)        & 5778  \\
Stellar radius & $R_\star$ ($R_\odot$)     & 1.0   \\
Orbital distance & $d_{\rm P}$ (AU)        & 1.523 \\
Planet mass    & $M_{\rm P}$ ($M_\oplus$)  & 0.107 \\
Planet radius  & $R_{\rm P}$ ($R_\oplus$)  & 0.531 \\
Zenith angle   & $\varphi$ (degrees)       & 20,45    \\
\hline
\end{tabular}
\flushleft
\label{ttone}
\end{table}

\begin{figure}
\centering
\vspace{-0.5cm}
\includegraphics[width=8.5cm]{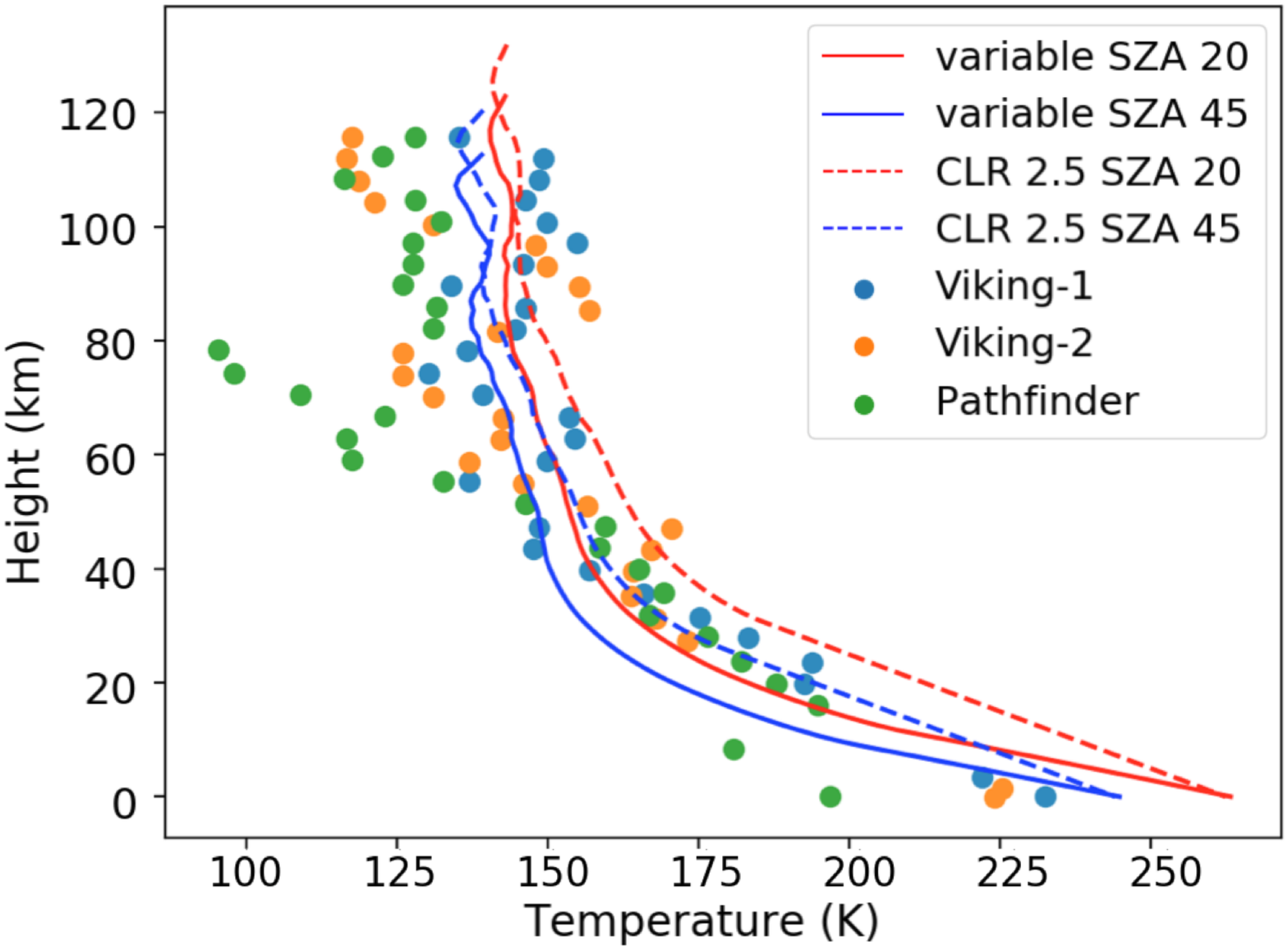}
\vspace{-0.5cm}
\caption{Synthetic (solid and dashed lines) and retrieved (dots) thermal profile of Mars. Red ($\varphi = 20^\circ$) and blue ($\varphi = 45^\circ$) lines: solid, variable CLR from equation~(\ref{gamma}); dashed, constant CLR, $\Gamma_{\rm c} = 2.5$~K~km$^{-1}$. Blue dots: \emph{Viking-1}; orange dots: \emph{Viking-2}; green dots: \emph{Pathfinder}. Data from \emph{Viking-1} and \emph{Viking-2} are taken in \citet{N76}, while the ones from \emph{Pathfinder} in \citet{M99}.}
\label{ftwo}
\end{figure}

We start from an initial isothermal profile at the  temperature $T (z) = T_{\rm s} = 180$~K. Using a SZA $\varphi = 45^\circ$ (approximately the \emph{Viking-2} landing site latitude), we find results consistent with data acquired during the descent of some of the major martian landers for a constant CLR, $\Gamma_{\rm c} = 2.5$~K~km$^{-1}$ (see Figure~\ref{ftwo}). This value is significantly lower than the dry adiabatic lapse rate, due to heating caused by suspended dust particle absorption, as well as circulation phenomena. The $P-T$ profile obtained using a variable with altitude CLR, computed through equation~(\ref{gamma}) is also shown. The {surface} temperatures differ of a few degrees. The atmosphere is globally colder when using the variable CLR.

\begin{figure}
\centering
\includegraphics[width=8cm]{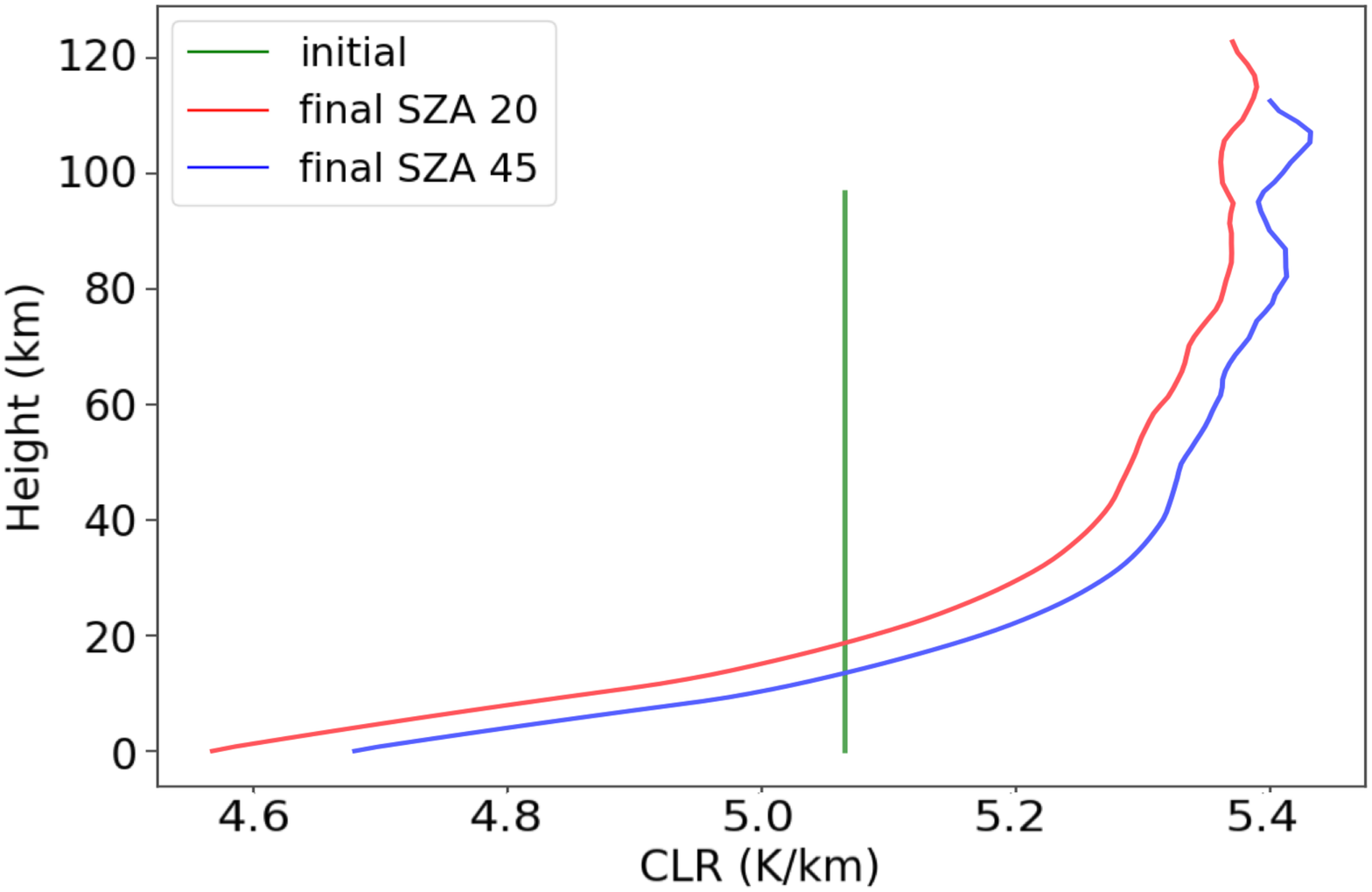}
\caption{CLR as a function of the altitude. Initial value: green vertical line; final profiles: $\varphi = 20^\circ$, red line; $\varphi = 45^\circ$, blue line.}
\label{fthree}
\end{figure}

In Figure~\ref{ftwo} are also shown two additional profiles, both of them derived using a SZA $\varphi = 20^\circ$, similar to \emph{Pathfinder} and \emph{Viking-1} landing sites' latitudes. The illumination angle plays an important role, e.g., in determining the surface temperature (around 20~K difference). The case in which the CLR is derived through equation (\ref{gamma}), $\Gamma_{\rm c} \sim 5$ produces a $P-T$ profile close to the one obtained using a constant CLR, $\Gamma_{\rm c} = 2.5$~K~km$^{-1}$ and SZA, $\varphi = 45^\circ$, although deviates significantly from this latter close to the {surface}, favouring a much warmer surface temperature, $\sim 265$~K. The evolution of the CLR is shown in Figure~\ref{fthree}. In the convection-dominated zone we obtain $\Gamma_{\rm c} \la 5$. 

\section{Results} \label{res}
We have constructed a grid of $6250$ models of dry planetary atmospheres considering a range of possible stellar and planetary parameters (see Table~\ref{ttwo}). For all models, the SZA is $\varphi = 60^\circ$, the emissivity parameter at TOA $\epsilon=0$, and the surface reflectivity $R_{\rm S} = 0.3$. A point grid in our parameter space is defined by two stellar characteristics,  effective temperature ($T_\star$) and incident flux at TOA ($S_\star$), and three planetary features, gravity ($g_{\rm P}$), {surface} pressure ($P_{\rm surf}$) and \ce{CO2} VMR ($\omega_{\ce{CO2}}$). Among the outputs, there are the emergent spectrum, either the planet-star flux ratio or the planet flux alone, and pressure-temperature $P-T$ profiles. 

In the light of the huge number of products, we deal with the general properties of derived synthetic atmospheres, such as e.g., the conditions for gas condensation, the relation of the planetary albedo with the stellar temperature and illumination, and the connection between {surface} temperature and orbital distance, stellar temperature, and stellar irradiation. 

Among the explored models, in a few cases gas temperatures are, at least in one layer, beyond the temperature range of our opacity table (see Table~\ref{kdisttab}). These models are tagged as \textit{crashed}, and included only in the characterization of \ce{CO2} condensation for statistic purposes.  
\begin{table}
\caption{Input parameters considered for the computation of the atmospheric grid.}
\hspace{-1cm}
\begin{tabular}{lcc}
\hline
Parameter                           & models & Values \\ \hline 
$S_\star$ at TOA (Wm$^{-2}$) & 5 & $\Delta S_\star = 270$, $120 - 1200$ \\ 
$T_\star$ (K) & 5 & $\Delta T_\star = 1000$, $3000 - 7000$ \\ 
$g_{\rm P}$ (m/s$^2$)  & 5 & $\Delta g_{\rm P} = 11.25$, $5 - 50$\\
$P_{\rm surf}$ (bar)  & 10           & log$_{10}$($\Delta P/{\rm bar}$) = $\sim0.22$, $0.3 - 30$ \\
$\omega_{\ce{CO_2}}$   & 5           & 0.0001, 0.01, 0.1, 0.5, 0.9
\\
\hline
\end{tabular}
\flushleft
\label{ttwo}
\end{table}

\subsection{\ce{CO2} condensation}
\begin{figure*}
\centering
\includegraphics[width=16cm]{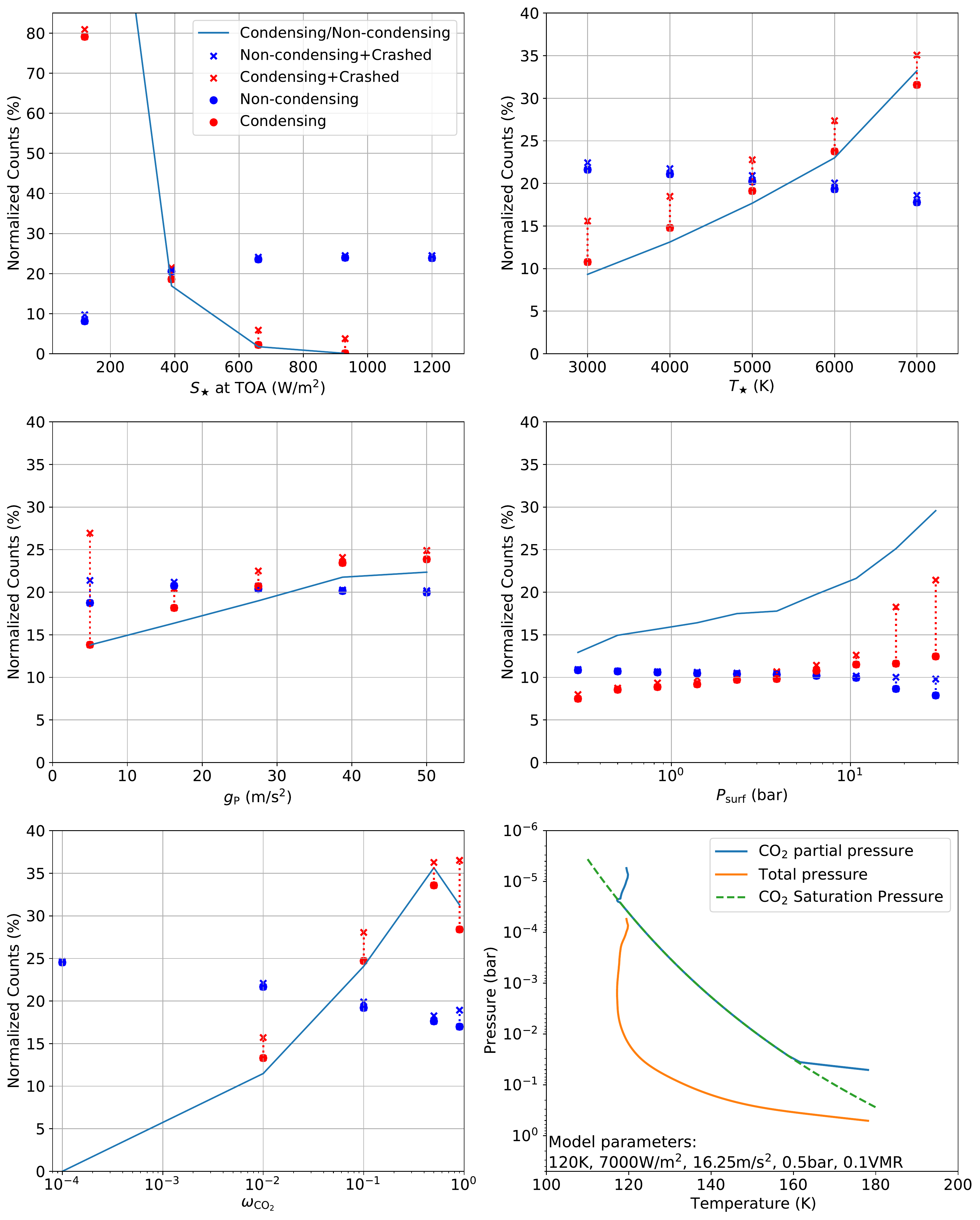}
\caption{The distribution of  model parameters in the case of condensing (red filled circles) and non-condensing (blue filled circles) models. Both classes of models are normalized to their total counts. A dashed line, bounded by crosses, marks the variation introduced by \textit{crashed} models (see text). The blue line describes the variation in the ratio between the number of condensing and non-condensing models. In the bottom right panel is shown the  \ce{CO2} partial pressure profile as a function of temperature for a condensing atmosphere (solid blue line); the green dashed line represents the \ce{CO2} phase diagram, while the orange solid line the total pressure profile; modelling parameters are reported as labels.}
\label{ffour}
\end{figure*}

In Figure~\ref{ffour}, we present the distribution of parameters against the number of both condensing ($\sim1000$) and non-condensing ($\sim 5000$) models. In the figure we also show the effect of including crashed ($\sim 250$) models and the ratio of condensing and non-condensing models. Since upon convergence such models might end up either condensing and non-condensing \ce{CO2}, we add them to both classes of models.

Looking at the scatter in the parameter distributions, it is clear that not all of them play equivalent roles in characterizing the condensation process, with stellar parameters being (relatively) more influential than planetary ones. The number of condensing models dramatically decreases with increasing incident stellar flux: atmosphere with high illumination gets hotter than the lower flux counterpart, thus exhibiting higher saturation pressures of \ce{CO2}. Most of the condensing models occur under an irradiation $S_\star \la 400$~Wm$^{-2}$. For similar considerations the number of condensing models increases together with the stellar temperature: atmospheres illuminated by cold stars absorb more radiation and get hotter than atmospheres of planets orbiting hotter stars. 

The condensation occurrence is largely insensitive to {surface} pressure, as long as we do not consider crashed models. Low gravity models tend to prevent condensation, although in general non condensing models are evenly distributed among all the assumed values of surface gravity, pointing out the scarce importance of this parameter. This is particularly evident if we ascribe crashed models to the non-condensing atmosphere set. The \ce{CO2} mixing ratio exhibits wider variations in condensing cases because \ce{CO2} partial pressure is evidently related to condensation. 

Finally, in Figure \ref{ffour} (bottom right panel) we show a typical $P-T$ profile for a condensing model. Anytime at a location $z$ in the atmosphere (with local $P_z$ and $T_z$), the \ce{CO2} partial pressure $P_{\rm \ce{CO2}} (z)$ exceeds the local vapor saturation pressure $P^{\rm sat}_{\rm \ce{CO2}}(T_z)$ we set $P_{\rm \ce{CO2}} (z) = P^{\rm sat}_{\rm \ce{CO2}}(T_z)$. In other words, we implicitly suppose that latent heat released by \ce{CO2} condensation warms the layer where condensation occurs, and alters the local thermal profile.

\subsection{Albedo}\label{albedo}
\begin{figure}
\centering
\includegraphics[width=8cm]{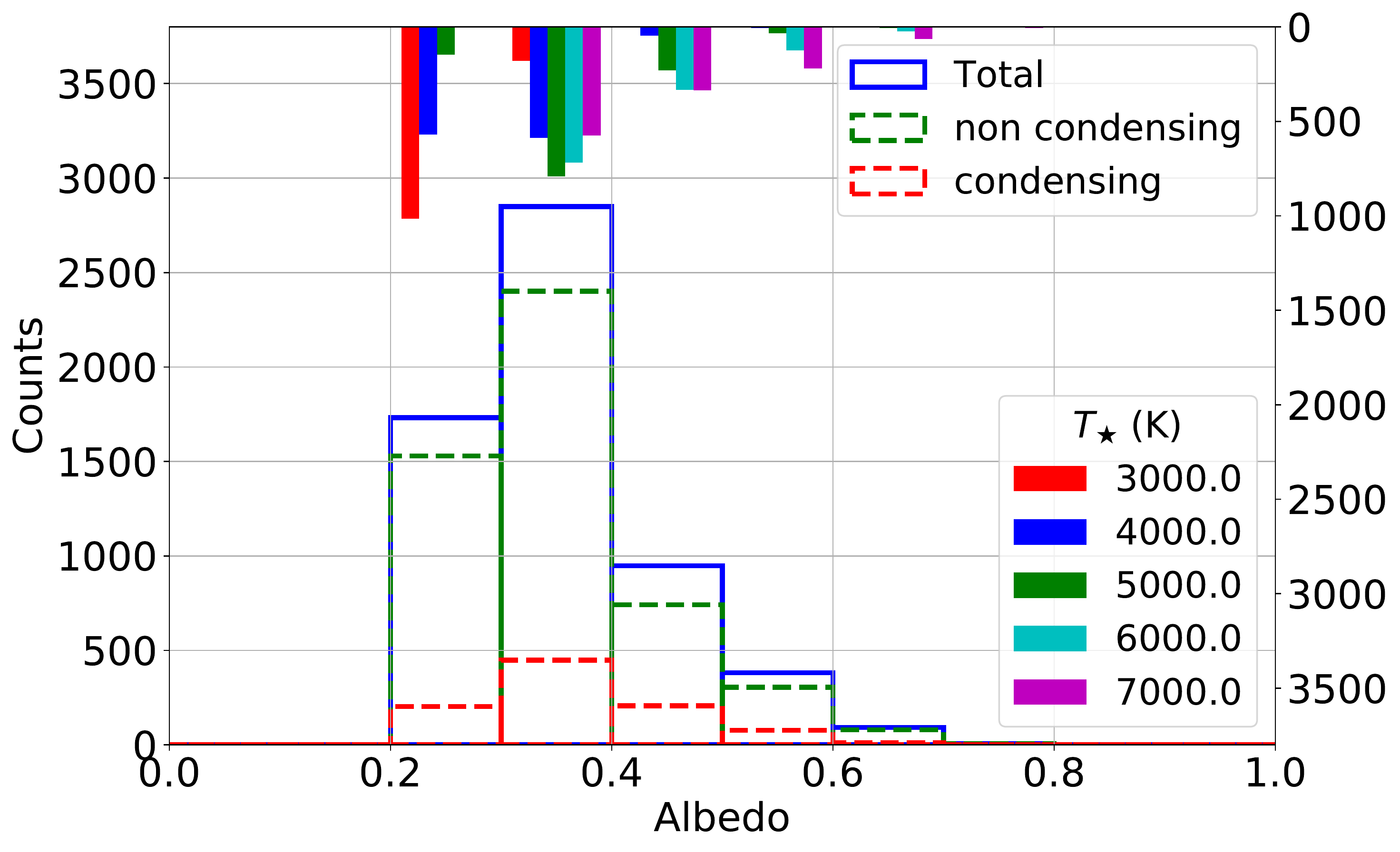}
\caption{Histogram (counts on the left ordinate axis) of modelled albedos for condensing (red dashed), and non-condensing (green dashed) models; the solid blue line describes the whole set of models. For each albedo bin, the stellar temperature ($T_\star$) distribution is shown as a reversed histogram whose counts are displayed in the right ordinate axis.} 
\label{ffive}
\end{figure}

In Figure \ref{ffive} we present the modelled planetary albedo obtained by the ratio between outgoing and incoming solar radiation fluxes. The albedo ranges from 0.2 to 0.7, with the majority of models  ($\sim4500$) lying in the range $0.2-0.4$, a value close to the reflectivity of the surface. In a smaller fraction of models ($\sim1500$) the albedo ranges over 0.4. The albedo depends strongly on the stellar temperature because of the wavelength dependence of the Rayleigh scattering, equation (\ref{eq:SR}): possessing a wider infrared spectral distribution, the irradiation from colder stars is less scattered than the radiation emitted from hotter stars. The highest albedo values are obtained for stellar temperatures over $6000$~K.

\subsection{A guess on habitability}
\begin{figure}
\centering
\includegraphics[width=8cm]{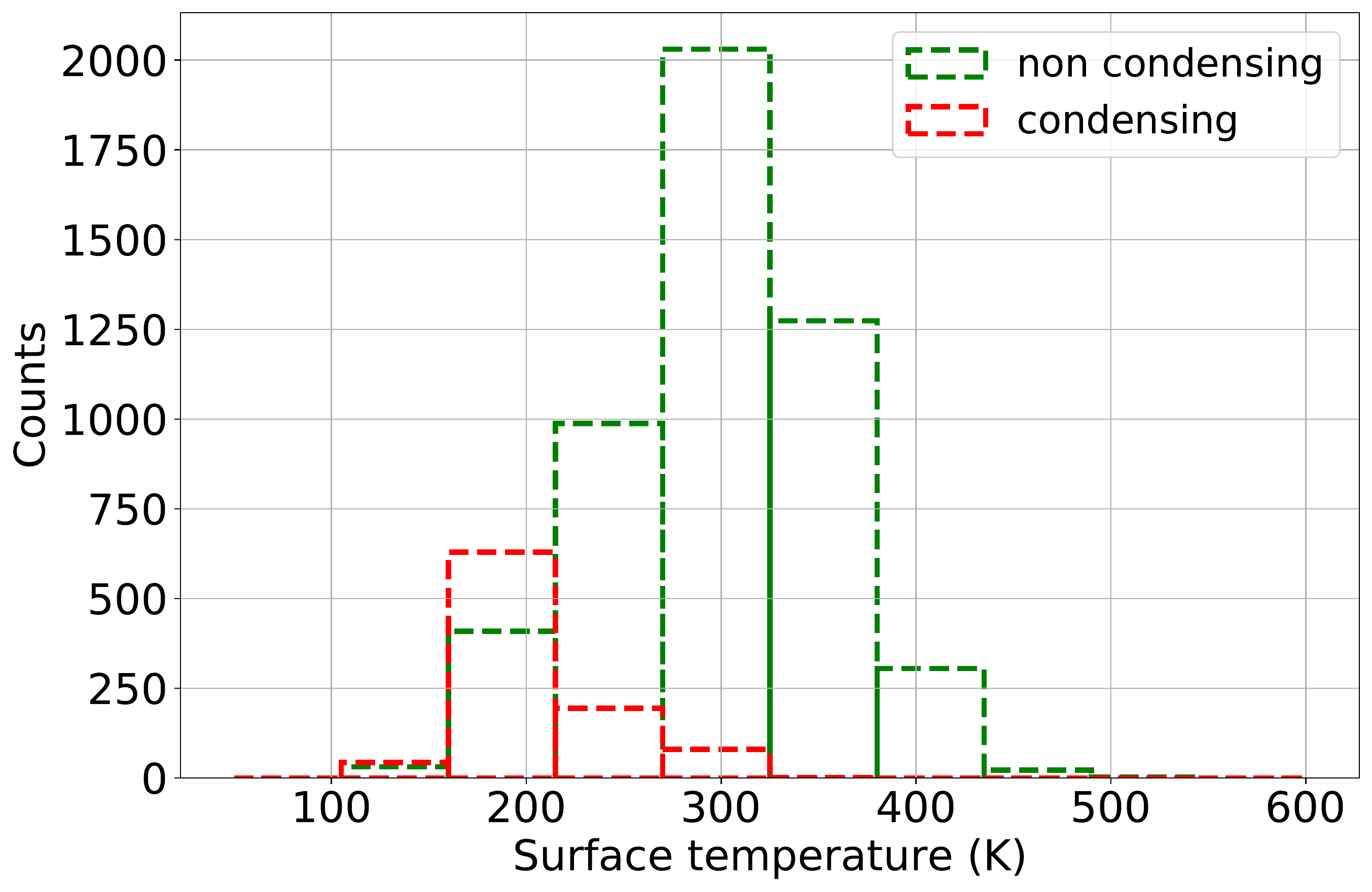}
\caption{Histogram of modelled {surface} temperatures for condensing (red dashed), and non-condensing (green dashed) models.}
\label{fsix}
\end{figure}

In this section, we use the {surface} temperature as a proxy for habitability. We don't discuss here what makes an extraterrestrial world habitable (see e.g., \citealt{MB18}). Since the atmospheric pressure affects the liquid water temperature range that is commonly used to define planetary habitability, for the sake of simplicity we term as \emph{habitable} those models in which the {surface} temperature is between the freezing and boiling points of water. The distributions of {surface} temperatures obtained for condensing and non-condensing models are different (Figure~\ref{fsix}), with the former confined to the lowest temperatures, peaking at 200~K, while the latter presenting a broader distribution centered at 300~K. Taking into account constraints posed by the phase diagram of water we derive that $\sim 3500$ models would have liquid water on the surface, and would be, potentially, habitable. Since water is essential to life as we know it, this should be the first step for narrowing down which requirements are needed. On the other hand, this very first bit of information neglects other demands of life, such as e.g., a source of carbon, an energy source, and essential nutrients. Moreover, it could be misleading being too Earth-centric. 

\begin{figure}
\centering
\includegraphics[width=8cm]{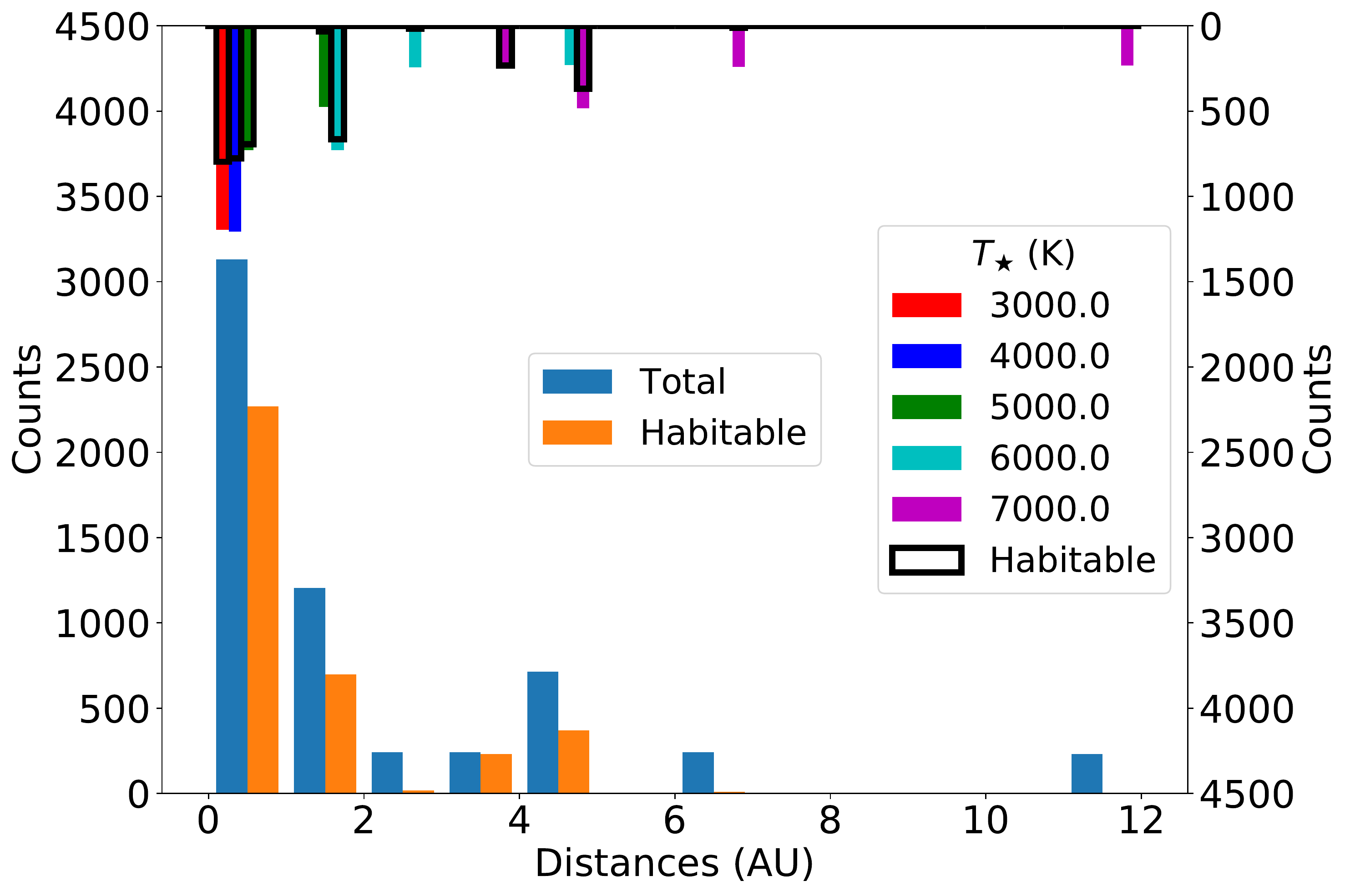}
\caption{Histogram of modelled distances (left and bottom axis), for the totality of the models (blue), and for the subset of the habitable planets (orange), together with the distribution of the stellar temperatures ($T_\star$, right and top axis). In this latter diagram the black empty boxes mark the fraction of habitable planets.}
\label{fseven}
\end{figure}

The connection between {surface} temperature and orbital distance, stellar temperature and illumination is presented in Figure \ref{fseven}. About two thirds of our modelled planets lie at distances lower than 2~AU, while the remaining $\sim 2000$ are at farthest locations from their central stars, with some of them having very large orbital distances, up to 12~AU. Given the assumed flux range, we do not include planets orbiting hot stars ($T_\star \ga 6000$~K) at distances lower than 1~AU, while those planets orbiting colder stars ($T_\star \la 4000$~K) are never beyond 2~AU. For orbital distances $d_{\rm P} \la 1$~AU, in about two thirds of cases the planets are habitable and orbit around stars whose temperatures are $T_\star \la 5000$~AU. In the second bin, comprising planets orbiting between 1 and 2~AU, we get habitable conditions only for stars with $T_\star = 5000$ and 6000~K. The rest of the sample models needs hot stars to be habitable. Farther than $d_{\rm P} = 6$~AU, there are no more habitable models.

\begin{figure*}
\centering
\includegraphics[width=18cm]{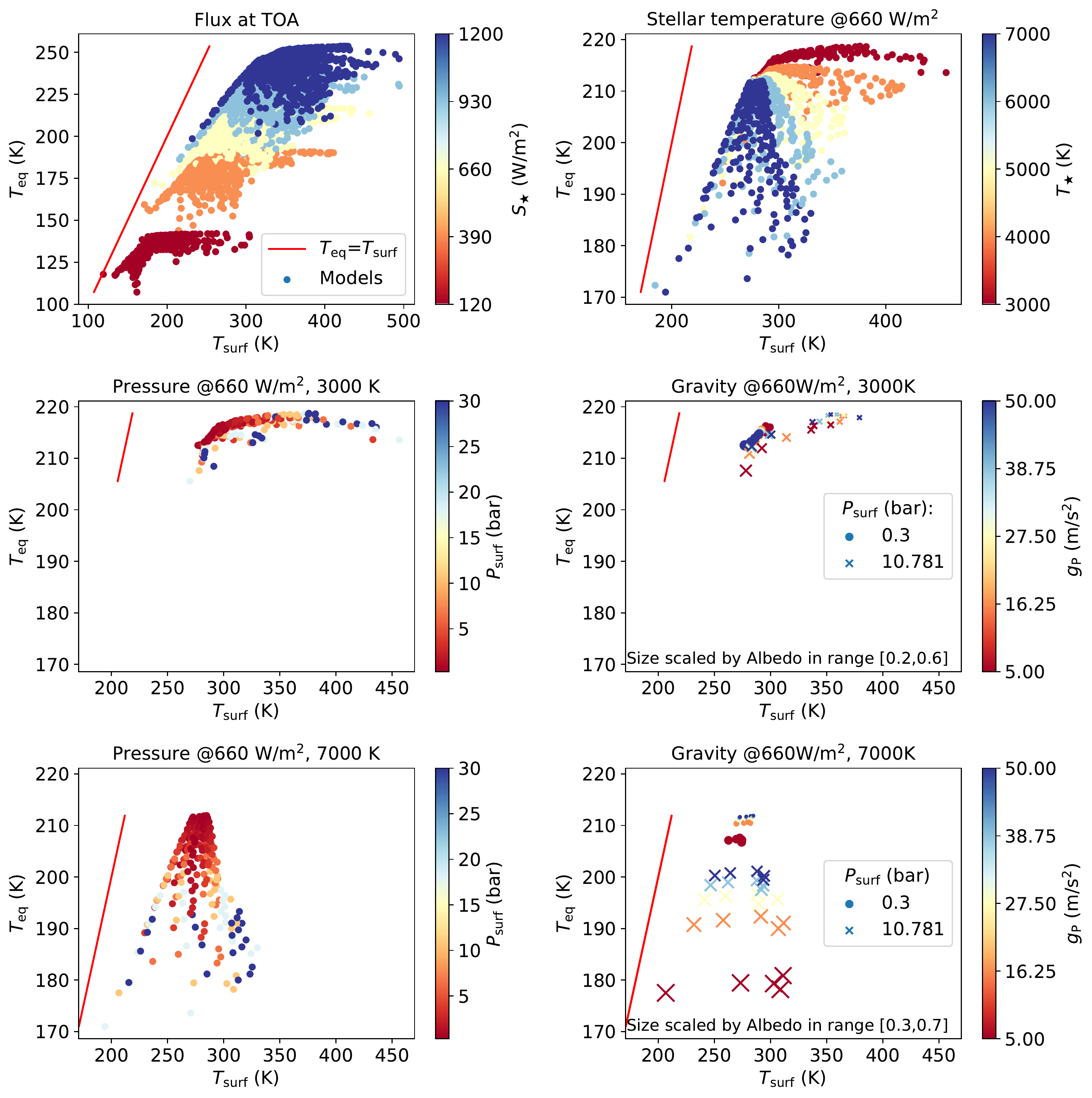}
\caption{Scatter of modelled {surface} temperatures, $T_{\rm surf}$ versus the estimated equilibrium temperatures, $T_{\rm eq}$, computed using equation (\ref{teq}) and the derived albedo. The red line represents the loci defined by the equivalence $T_{\rm eq} = T_{\rm surf}$. Top left panel: the whole set of models labelled by the incident flux, $S_\star$; top right panel: selected models with $S_\star = 660$~W~m$^{-2}$, labelled by stellar temperature; middle panels: selected models with $S_\star = 660$~W~m$^{-2}$ and $T_\star = 3000$~K, labelled by {surface} pressure (left) and gravity (right, for two values of the {surface} pressure); bottom panels: the same as in middle panels for $T_\star = 7000$~K. In the middle and bottom right panels, the sizes of symbols are related to the values of the planetary albedo.}
\label{feight}
\end{figure*}

\subsection{{Surface} and planetary equilibrium temperatures}

In Figure~\ref{feight} we present the relation between the modelled {surface} temperature and the equilibrium temperature, derived using equation (\ref{teq}) in which the albedo is a result of the modelling procedure. We firstly present the entire set of models (in the top left panel), with each model labelled by the stellar flux incident at the TOA, $S_\star$. In order to understand how the temperatures are related to the input parameters, we extract subsets of data points, defined by selecting, consecutively, specific values of different model parameters, namely the flux at TOA (top right panel, labelled by stellar temperature, $T_\star$), two stellar temperatures (mid and bottom left panels, labelled by {surface} pressure, $P_{\rm surf}$) and two {surface} pressures (mid and bottom right panels, labelled by gravity, $g_{\rm P}$). We also highlight albedo values (mid and bottom right panels) through the sizes of the symbols exploited to flag a model in the temperatures' plane.

By looking at the entire set (top left panel), it is evident that {surface} temperatures are always greater than the equilibrium temperatures, and both of them increase with the incident flux, as the flux enters directly in the estimation of planetary temperatures. Selecting an intermediate value for the flux, $S_\star = 660$~W~m$^{-2}$, we may analyse how different parameters shape its dispersion properties and therefore how much the {surface} temperature differs from the equilibrium temperature. We find that different stellar types have completely different dispersion (top right panel). Cold stars (red circles) provide very similar equilibrium temperatures with a dispersion $\Delta T_{\rm eq} \sim 20$~K, while hotter stars (blue circles) show broader temperature distributions increasing with the stellar temperature up to $\sim 50$~K. The different equilibrium temperatures between cold and hot stars are primarily due to the very different modelled albedo (see Figure~\ref{ffive}). Planets facing cold stars have a small range of possible albedos close to the reflectivity of the surface ($R_{\rm S} =0.3$) while the ones facing hot stars showing a much broader range of albedos (up to 0.7). To understand the dispersion of the {surface} temperatures, in the last (mid and bottom) panels we show the effects of the {surface} pressure, gravity and albedo in the models produced using the end points of the assumed stellar temperature distribution, i.e. $T_\star = 3000$ (mid panels) and 7000~K (bottom panels). 

In both cases, the dispersion in the {surface} temperatures increases with the {surface} pressure from $\Delta T_{\rm surf} \sim 20$~K at 0.3~bar to $\Delta T_{\rm surf} \sim120$~K at 30~bar. The role of gravity is marginal, and it contributes only partially in hot star environments by lowering the dispersion. The opposite is true for the albedo (cf. the symbol sizes in the two lowest right panels). {In general, the highest surface temperatures are reached for large values of the surface pressure. This behaviour is driven by the \ce{CO2} greenhouse effect. As the \ce{CO2} content increases and the atmosphere becomes opaque to outgoing infrared radiation, this determines the maximum greenhouse effect (see \citealt{Kastetal93} and \citealt{Kop13}). This is evident in the bottom left panel of Figure~\ref{feight}, where the increase in the surface pressure does not drive anymore the increase in the surface temperature. The influence of the incident stellar spectrum, can be seen in the top right panel where cooler stars show a less efficient Rayleigh scattering, resulting in higher surface temperatures.}

\begin{figure*}
\centering
\includegraphics[width=15cm]{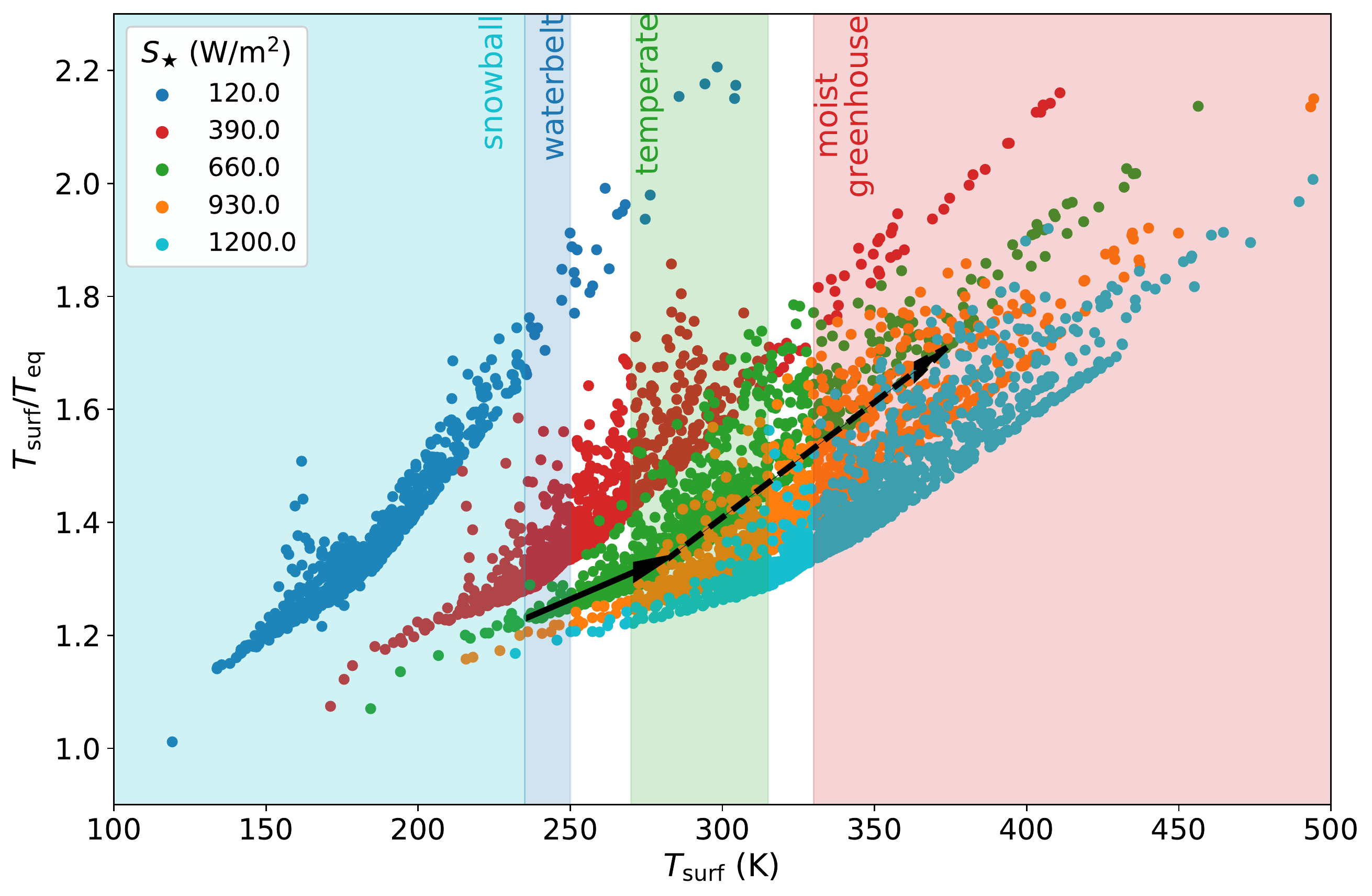}
\caption{{Surface} temperatures versus the ratio $T_{\rm surf}/T_{\rm eq}$. Colors indicate different levels of incident stellar flux. Two arrows mark temperature variations obtained by varying, the stellar temperature (solid black) and the \ce{CO2} mixing ratio (dashed black), setting all the other parameters to the values $S_\star=660$~W~m$^{-2}$, $P_{\rm surf}=30$~bar, $g_{\rm P}=50$~m~s$^{-2}$. The solid line represents $T_\star = 7000 \to 3000$~K, with $\omega_{\ce{CO2}} = 1 \times 10^{-4}$, while the dashed line connects the model with $T_\star = 3000$~K and $\omega_{\ce{CO2}} = 1 \times 10^{-4}$, to the one with $T_\star = 3000$~K and $\omega_{\ce{CO2}} = 0.9$. For illustration only, we show on the background the different climate regimes identified by \citet{W17}.}
\label{fnine}
\end{figure*}

\section{Conclusions}
The wealth of super-Earth detections suggests that terrestrial planets may be abundant in our galaxy. A large fraction of the stars are likely to harbour rocky planets, and the available observations suggest a heterogeneity of climates. Such variety  depends on the planetary physical characteristics (e.g., stellar illumination, orbit, and size) and the composition of its atmosphere. Among others, radiative transport of stellar and thermal radiation through gas and aerosols is a primary key process in a climate system. 

In this work, we construct and exploit a 1D RT code based on the discrete-ordinates method in plane-parallel geometry. The RT modelling results are linked to a convective flux that redistributes energy at any altitude producing atmospheric profiles in RC equilibrium. Here, we present simulations of dry atmospheres in terrestrial-like planets with varying concentration of \ce{CO2}, around main sequence stars (from M to F types), over a wide range of stellar fluxes and temperatures.

We initially probe the effect of input parameters on the onset of \ce{CO2} condensation. As expected, the (average) stellar illumination and \ce{CO2} mixing ratio are critical aspects in defining the possibility of condensation. Condensation is suppressed either when the atmosphere receives more than 400 W~m$^{-2}$, and with \ce{CO2} mixing ratios lower than 1\%. All other parameters affect the condensation process to a lesser extent.

The modelled albedo is close to the reflectivity of the surface for planets embedded in cold star ($\la 4000$~K) environments. The range of the albedo values extends if the planet is located around a hotter star ($\ga 5000$~K). Therefore, in the case of our simple atmospheric composition and geometric configuration the albedo acts as a proxy of reflectivity properties of the planetary surface. We recall that we consider a perfectly Lambertian surface, for which the hemispherical reflectance is simply equal to the reflectance factor independently of viewing geometry.

We also explore the conditions for (a restricted definition of) habitability both in terms of {surface} temperature to sustain liquid water and orbital distances. We find that more than half of the planets would be habitable, with the majority having orbital distances $\la  2$~AU and a small, but not negligible fraction extending up to 6~AU. 

Finally, we compare the equilibrium temperature of the atmosphere with the modelled {surface} temperature. The equilibrium temperature represents a sort of lower limit to the {surface} temperatures, with the extent of the deviation between the two temperatures driven by the characteristics of specific models. The equilibrium temperature is generally used to categorize the potential habitability of exoplanets with Earth-like planetary albedo of 0.3 or 0 (e.g., \citealt{Ang16}). As a proxy, the equilibrium temperature has a intrinsecally poor reliability because albedos can be very different, and surface temperatures typically exceed equilibrium temperatures by the amount of the atmosphere's greenhouse effect. Thus, despite the limited resources, planets should be observed intensely to characterize their atmospheres. 

Recently, a few studies typically based on global climate models addressed the problem of the removal of degeneracies arising from relations among observables by estimating albedo, equilibrium temperature, and surface temperature of rocky exoplanets (e.g., \citealt{W17,DG19}, for a limited manifold of models. Although less accurate, not properly accounting for clouds, sea ice and atmospheric and ocean dynamics, 1D models are computationally inexpensive and allow to expand significantly the explored parameter space. In Figure \ref{fnine} we show the ratio $T_{\rm surf}/T_{\rm eq}$ versus $T_{\rm surf}$ for the whole set of converged models ($\sim 6000$). Leaving aside extreme configurations, this relation appears to be roughly linear, with a steepness controlled by the stellar illumination, whose increase tends to moderate the deviation from $T_{\rm eq}$ of {surface} temperatures. For illustrative purposes only we have shown in the background the four stable climate states defined by \citet{W17}, through  mutually exclusive global mean surface temperatures, whose boundaries are separated by abrupt climatic transitions. As an example, we have drawn a path travelled by a data point varying some parameters, while taking fixed $S_\star$ at intermediate illuminations, i.e. $S_\star=660$~W~m$^{-2}$. Planets with similar characteristics in {surface} pressure, $P_{\rm surf}=30$~bar, gravity, $g_{\rm P}=50$~m~s$^{-2}$, and \ce{CO2} VMR $\omega_{\ce{CO2}} = 1 \times 10^{-4}$, that are initially in the snowball regime when orbiting hot stars, transit through waterbelt and temperate zones if the stellar temperature decreases to $T_\star = 3000$~K. From this location, if the \ce{CO2} VMR increases up to 0.9, our test planet reaches the moist greenhouse state. This simple case illustrates how low dimensional models retain significant value by allowing multidimensional parameter sweeps with relative ease. 

In conclusion, we have constructed a horizontally homogeneous model of RC equilibrium to compute the global average of a planetary atmosphere. The model has been applied to a large number of closely dry synthetic atmospheres, and the resulting pressure and thermal profiles have been interpreted in terms of parameter variability. In future works, we shall extend the chemical inventory of the modelled atmospheres, and explore how the relative humidity or cloud distributions are maintained in such idealized systems. Moist convection is dominated by phase changes, and therefore may present peculiar and unfamiliar aspects, that may be usefully isolated in a simple geometry.
 
\section*{Acknowledgements}
We acknowledge support from ASI-INAF agreement 2018-22-HH.0 \emph{Partecipazione alla fase B1 della missione ARIEL}, the project PRIN-INAF 2016 \emph{The Cradle of Life - GENESIS-SKA (General Conditions in Early Planetary Systems for the rise of life with SKA)}, {and support by INAF/Frontiera through the "Progetti Premiali" funding scheme of the Italian Ministry of Education, University, and Research}. {E.A. acknowledges the financial support of the Swiss National Science Foundation.}

We also acknowledge the \emph{Accordo Quadro INAF-CINECA (2017)} {and SCAN, of the INAF-Osservatorio Astronomico di Palermo,} for the availability of high performance computing resources and support.

\appendix
\section{Line modifications and continuum description}
Carbon dioxide line profiles are observed to be sub-lorentzian away from the line center. We use the scaling factor suggested by \citet{Per89} beyond 3 cm$^{-1}$ from the line center. Moreover, to adjust for the asymmetry of the line profile we applied the normalized correction factor given in \citet{Vvh77}. The CO$_2$ continuum is computed using a semi-empirical model, which provides the foreign continuum from 1 to $1000~\mu$m or $0.1 -10000$~ cm$^{-1}$ \citep[{MT\_CKD continuum v2.5.2,}][]{Mla12}. The \ce{CO2} molecule also experiences collision induced absorption due to encounters with other \ce{CO2} molecules. We account for these effects using the theoretical results of \citet{Gru97} in the interval $33-1000~\mu$m ($10-300$~cm$^{-1}$) and interpolating the experimental results of \citet{Bar04} in the interval $5-9~\mu$m ($1100-2000$~cm$^{-1}$). To avoid an overestimate of the absorption, each \ce{CO2} line is calculated out to 25 cm$^{-1}$ from the line center, hence (as suggested by \citealt{Mla12}) we consider continuum all the absorption produced further. 

In the case of \ce{H2O} we follow the prescriptions of \citet{Hal09} and \citet{Mis12}. We compute a Voigt profile within 40 times the Doppler width from the line center, and a Van Vleck-Weisskopf profile out to 25~cm$^{-1}$ from the natural wavenumber of the transition $\nu_0$. Also for water, we use the continuum model given in \citet{Mla12}, which provides both foreign and self continuum from 0.5 to $1000~\mu$m (0.1 to 20000~cm$^{-1}$). Such a model derives H$_2$O line shapes by fitting the continuum in spectral regions where it is best constrained by measurements. The same line shape is extended to spectral regions where the continuum is not well constrained or simply never been measured.

Recently, \citet{K17} implemented a new water vapor continuum model \citep{PR11}, derived from laboratory measurements taken at temperatures appropriate for atmospheres expected near the inner edge of the habitable zone. Such model provides a continuum absorption stronger than in the model put forward by \citet{Mla12}. While this is not influential on the results of this work, we shall update the spectroscopic database in future releases of our model, {e.g., testing the recently released 3.3 version of the MT\_CKD continuum}. 

\bsp    
\label{lastpage}

\begin{thebibliography}{99}
\bibitem[Allen(1973)]{A73} Allen, C. 1973, Astrophysical Quantities (University of London: The Athlone Press)
\bibitem[Anglada-Escud{\'e} et al.(2016)]{Ang16} Anglada-Escud{\'e}, G., Amado, P. J., Barnes, J., et al. 2016, Nature, 536, 437 
\bibitem[Baranov et al.(2004)]{Bar04} Baranov, Y.I., Lafferty, W.J., \& Fraser, G.T., 2004, J. Mol. Spec., 228, 432
\bibitem[Bonfils et al.(2013)]{Bon13} Bonfils, X.; Delfosse, X., Udry, S., et al. 2013, \aap, 549, A109
\bibitem[Cox(2000)]{Cox00} Cox, A. N. 2000, Allen's astrophysical quantities (Springer-Verlag)
\bibitem[Del Genio et al.(2019)]{DG19} Del Genio, A. D., Kiang, N. Y., Way, M. J., et al. 2019, \apj, 884, 75
\bibitem[Fressin et al.(2013)]{F13} F. Fressin, G. Torres, D. Charbonneau et al. 2013, \apj, 766, 81
\bibitem[Gordon et al.(2017)]{G17} Gordon, I. E., Rothman, L. S., Hill, C., et al. 2017, JQSRT, 203, 3
\bibitem[Gruzska \& Borysow(1997)]{Gru97} Gruzska, M., \& Borysow, A., 1997, \icarus , 129, 172
\bibitem[Halevy et al.(2009)]{Hal09} Halevy, I., Pierrehumbert, R.T., \& Schrag, D.P. 2009, \jgr, 114, D18112 
\bibitem[Howard(2013)]{H13} Howard, A. W. 2013, Science, 340, 572
\bibitem[Hu \& Ding(2011)]{HD11} Hu, Y. \& Ding, F. 2011, \aap, 526, A135
\bibitem[Kasting, Whitmire \& Reynolds(1993)]{Kastetal93} Kasting J.~F., Whitmire D.~P., \& Reynolds R.~T., 1993, \icarus, 101, 108
\bibitem[Kopparapu et al.(2013)]{Kop13} Kopparapu, R.K., Ramirez, R., Kasting, J.F., et al. 2013, \apj, 765, 131
\bibitem[Kopparapu et al.(2016)]{K16} Kopparapu, R. K., Wolf, E. T., Haqq-Misra, J., et al. 2016, \apj, 819, 1
\bibitem[Kopparapu et al.(2017)]{K17} Kopparapu, R. K., Wolf, E. T., Arney, G., Batalha, N. E., Haqq-Misra, J., Grimm, S. L., \& Heng, K. 2017, \apj, 845,~5
\bibitem[Magalh{\~{a}}es et al.(1999)]{M99} Magalh{\~{a}}es, J. A., Schofield J. T. \& Seiff A., 1999, \jgr, 104, 8943
\bibitem[Manabe \& Strickler(1964)]{MS64} Manabe, S. \& Strickler R. F. 1964, J. Atm. Sci., 21, 361
\bibitem[Manabe \& Wetherald(1967)]{MW67} Manabe, S. \& Wetherald, R. T. 1967, J. Atm. Sci., 24, 241
\bibitem[\protect\citeauthoryear{Mayor, Lovis \& Santos}{2014}]{May14} Mayor M., Lovis C., Santos N.~C., 2014, Natur, 513, 328
\bibitem[Meadows \& Barnes(2018)]{MB18} Meadows V. S., \& Barnes, R. K. (2018), in Handbook of Exoplanets, Deeg, H., \& Belmonte, J. (eds), p. 54, Springer
\bibitem[Mischna et al.(2012)]{Mis12} Mischna, M.A., Lee, C., \& Richardson, M. 2012, \jgr, 117, E10009 
\bibitem[Mlawer et al.(2012)]{Mla12} Mlawer,, E., Payne, V., Moncet, J.L., Delamere, J., Alvarado, M., \& Tobin, D., 2012,  Phil. Trans. R. Soc., 370, 2520
\bibitem[Nier et al.(1976)]{N76} Nier, A. O., Hanson, W. B., Seiff, A., McElroy, M. B., Spencer, N. W., Duckett, R. J., Knight, T. C. D.,  \& Cook, W. S. 1976, Science, 193, 786
\bibitem[North \& Erukhimova(2009)]{N09} North, G. R. \& Erukhimova, T. L. 2009, Atmospheric Thermodynamics: Elementary Physics and Chemistry (Cambridge University Press)
\bibitem[Ozak, Aharonson \& Halevy(2016)]{O16} Ozak N., Aharonson O., \& Halevy I. 2016, \jgr E, 121, 965 
\bibitem[Paynter \& Ramaswamy(2011)]{PR11} Paynter, D. J., \& Ramaswamy, V. 2011, JGRD, 116, D20302
\bibitem[Perrin \& Hartmann(1989)]{Per89} Perrin, M.~Y., \& Hartmann, J.~M. 1989, \jqsrt, 42, 311 
\bibitem[Stamnes et al.(1988)]{Sta88} Stamnes, K., Tsay, S.-C., Jayaweera, K., \& Wiscombe, W. 1988, \ao, 27, 2502
\bibitem[Toon et al.(1989)]{Too89} Toon, O.B., McKay, C.P., Ackerman, T.P., \& Santhanam, K. 1989, \jgr, 94, 16287 
\bibitem[van Vleck \& Huber(1977)]{Vvh77} van Vleck, J.H., \& Huber, D.L. 1977, Rev. Mod. Phys., 49, 939
\bibitem[Vardavas \& Taylor(2007)]{VT07} Vardavas, I. M., \& Taylor, F. V. 2007, Radiation and Climate (Oxford University Press)
\bibitem[von Paris et al.(2010)]{vonParetal10} von Paris, P., Gebauer, S., Godolt, M., et al. 2010, \aap, 522, A23
\bibitem[Wolf \& Toon(2015)]{WT15} Wolf, E. T., \& Toon, O. B. 2015, \jgr~(Atm.), 120, 5775
\bibitem[Wolf et al.(2017)]{W17} Wolf, E. T., Shields, A. L., Kopparapu, R. K., Haqq-Misra, J. \&  Toon. O. B. 2017, \apj, 837, 107
\bibitem[Wordsworth et al.(2010)]{Woretal10} Wordsworth R.~D., Forget F., Selsis F., Madeleine J.-B., Millour E., \& Eymet V. 2010, \aap, 522, A22
\bibitem[Yang et al.(2016)]{Y16} Yang, J., Leconte, J., Wolf, E. T. 2016, \apj, 826, 222
\end{thebibliography}
\end{document}